\documentclass[prl,twocolumn,showpacs]{revtex4-1}
\usepackage{ae}
\usepackage{graphicx}
\usepackage{amssymb}
\usepackage{amsmath}
\usepackage{amsthm}
\usepackage{sidecap}
\usepackage{wasysym}
\usepackage{pifont}
\newcommand{\deltasp}{\ensuremath{\delta_{\mathrm{sp}}}}
\newcommand{\mudiff}{\ensuremath{\mu_{\mathrm{diff}}}}
\newcommand{\HIWI}{\ensuremath{\hat{H}_{\mathrm{IWI}}}}
\newcommand{\HIWT}{\ensuremath{\hat{H}_{\mathrm{IWT}}}}
\newcommand{\eV}{\ensuremath{\mathrm{eV}}}
\newcommand{\meV}{\ensuremath{\mathrm{meV}}}
\newcommand{\nm}{\ensuremath{\mathrm{nm}}}
\newcommand{\K}{\ensuremath{\mathrm{K}}}
\newcommand{\T}{\ensuremath{\mathrm{T}}}
\newcommand{\Cex}{\ensuremath{\mathcal{C}_{\mathrm{ex}}}}
\newcommand{\vx}{\ensuremath{\mathbf{x}}}
\newcommand{\epssp}{\ensuremath{\varepsilon_{\mathrm{sp}}}}
\newcommand{\epssub}{\ensuremath{\varepsilon_{\mathrm{sub}}}}
\newcommand{\dIW}{\ensuremath{d_{\mathrm{IW}}}}

\begin{document}
\title{True bilayer exciton condensate of one-dimensional electrons}

\author{A.~Kantian}
\author{D.~S.~L.~Abergel}
\affiliation{Nordita, KTH Royal Institute of Technology and Stockholm
University, Roslagstullsbacken 23, SE-106 91 Stockholm, Sweden}

\begin{abstract} 
	We theoretically predict that a true bilayer exciton condensate,
	characterised by off-diagonal long range order and global phase
	coherence, can be created in one-dimensional solid state electron
	systems.
	The mechanism by which this happens is to introduce a single
	particle hybridization of electron and hole populations, which locks
	the phase of the relevant mode
	and hence invalidates the Mermin--Wagner theorem.
	Electron--hole interactions then amplify this tendency towards
	off-diagonal long range order, enhancing the condensate properties
	by more than an order of magnitude over the noninteracting limit.  
	We show that the temperatures below which a substantial condensate
	fraction would form could reach hundreds of Kelvin, a benefit of the
	weak screening in one-dimensional systems.
\end{abstract}
\date{\today}
\maketitle

Excitons are composite bosons formed from paired electrons and holes.
They can be produced either by optical pumping of carriers between bands, 
or by bringing physically separate electrons and holes in close proximity.
This latter type, called ``\textit{bilayer excitons}'' occur because the
mutual Coulomb interaction between the layers induces a many body
instability, allowing the excitonic state to form.
Under certain conditions, these composite bosons may condense into a
Bose--Einstein condensate with off-diagonal long range order
(ODLRO) and a global coherent phase~\cite{BookMoskalenko2000}.
Such a condensate has been observed for optically pumped excitons
\cite{Snoke2002}, and bilayer excitons in the quantum Hall regime
\cite{Eisenstein-Nature432}. 
In zero magnetic field, exciton based generation of thermoelectricity
has been proposed~\cite{Wu-PRAppl2}, and when condensed,
bilayer excitons have been predicted to
provide electrical transport across their bulk that is only limited by
contacts and a linking resistor~\cite{Su2008}. The predicted
dissipationless current between layers is a direct result of the
existence of the condensate and has been explored as the basis of
valuable devices, such as ultra low power
transistors~\cite{Banerjee-EDL30}.

However, a condensate of bilayer excitons in zero magnetic field has
never been observed in an experiment on two-dimensional materials
\cite{Pillarisetty-PRL89, Seamons-PRL102, Kim-PRB83, Gorbachev-NatPhys8,
Gamucci-NatComms5}.
Possible reasons include the critical temperature of the many body
instability being simply too low, due to strong screening of the
interlayer Coulomb interaction in two dimensions \cite{Sodemann-PRB85}. 
Another reason might be the destruction of Fermi surface nesting by charged impurity disorder 
\cite{Abergel-PRB86, Abergel-PRB88}.
Both of these factors could be mitigated by working with one-dimensional (1D) bilayers,
such as two parallel nanowires \cite{Abergel-APL106}.
Screening is known to be generally much weaker in 1D
systems~\cite{Schulz1993a,Meyer2009}, implying that the interlayer
interaction would be more effective in 1D.
As robustness against disorder derives from the magnitude of the order 
parameter \cite{Abergel-PRB86, Abergel-PRB88} this absence of screening
would enhance the stability of the 1D exciton condensate (EC) in this
respect as well. 

The chief obstacle to any condensate of quantum particles in 1D is the
Mermin--Wagner (MW) theorem, which prohibits spontaneous breaking
of a continuous symmetry, and thus ODLRO, due to the enhancement of quantum fluctuations
\cite{BookGiamarchi2003}.
In this work, we show that for 1D bilayer excitons, a very weak single
particle tunneling between the two layers can lead to a true EC with
ODLRO as the tunneling {\it explicitly} locks the phase of the relevant
mode and thus the MW theorem no longer applies.
Electron--hole attractions can then strongly feed into this small
tendency towards ODLRO, resulting in large enhancements of all
properties of the EC. 
This EC is a true many body condensate characterized by one large and
one small excitation gap, both of which can be probed experimentally. 

We employ highly accurate density matrix renormalization group (DMRG)
numerics~\cite{Schollwock2011} to compute the ground and thermal
state of the many body system. 
We show that the smaller gap sets the temperature scale on which
crossover to the EC occurs.
We also describe experimental probes of the EC by determining the
nonlinear DC current--voltage characteristic of an interlayer transport
measurement, and computing the density of states that would be probed in
an STM experiment.
Finally, we compute the ground states for systems with realistic length
and energy scales and show that the EC can be realized at high
temperatures after accounting for long range electron--electron
interactions.

\begin{figure}[t]
\centering
\includegraphics[width=1\columnwidth,trim = 45mm 30mm 50mm 23mm, clip]
{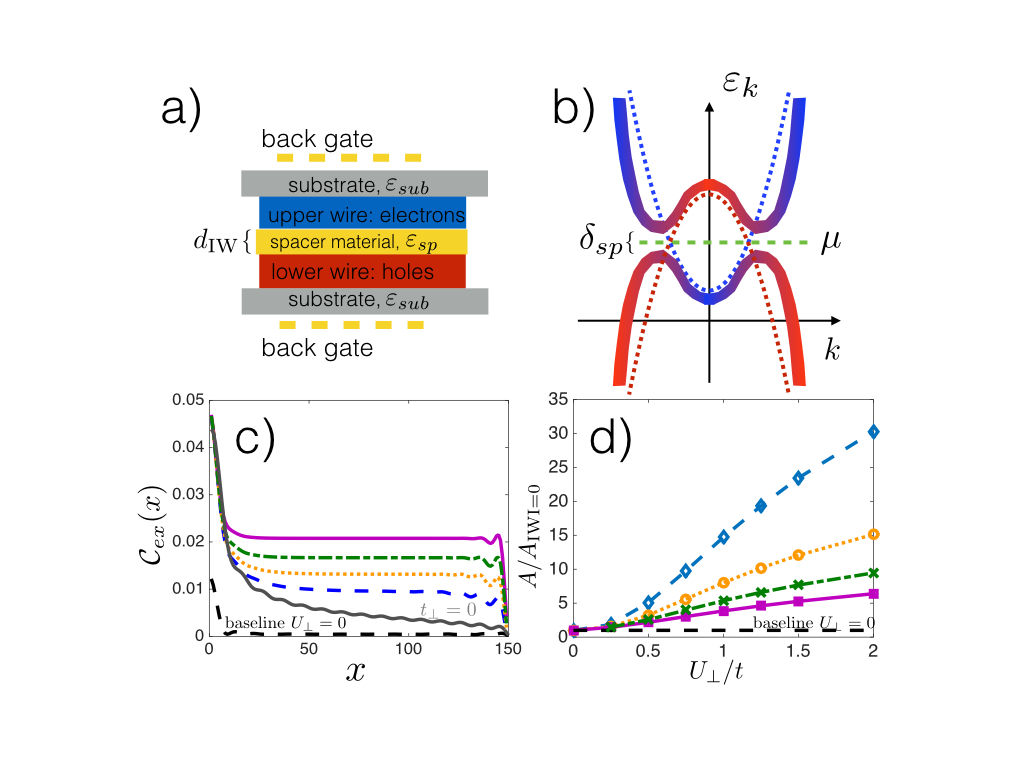}
\caption{(a) Sketch of the proposed system.
(b) Band structure of noninteracting electrons in parallel 
nanowires with weak interwire tunneling $t_\perp$.
The tunneling forces a gap $\deltasp = 2t_\perp$ (solid shaded lines)
to open at the Fermi level.
(c) Spatial dependence of the exciton--exciton correlator, showing
the strong enhancement of excitonic off-diagonal long range order
in the ideal model at zero temperature when $U_\perp=2t$, for
$t_\perp=0.001t$ (blue dashed line),
$t_\perp=0.0025t$ (orange dotted), 
$t_\perp=0.005t$ (green dash-dotted),
$t_\perp=0.01t$ (purple solid), 
$t_\perp=0$ (grey solid).
Free electrons ($U_\perp=0$) with $t_\perp=0.01$ (black dashed) shown for
comparison.
(d) Ratio of the order parameter $A$ of the exciton condensate with
interactions ($U_\perp\neq 0$) to noninteracting case (i.e. free fermions,
$U_\perp=0$). In all cases, we see that sufficient $U_\perp$ can
enhance the excitonic order by an order of magnitude or more.
The line styles match (c). Results in (c) and (d) are for $M=300$.}
\label{fig:Fig1}
\end{figure}

We consider a generic setup, two parallel quasi-1D electron systems
(``\textit{wires}'', hence), shown in 
Fig.~\ref{fig:Fig1}(a). 
Gates shift the electron bands such that the minimum of the conduction
band for the upper wire is below the maximum of the valence band for
the hole like lower wire.
Weak interwire (IW) tunneling $t_\perp$ results in a joint chemical
potential and, in the absence of interactions, the opening of a small
single particle gap $\deltasp=2t_\perp$
(Fig.~\ref{fig:Fig1}(b)). 
To be compatible with DMRG, we consider a 1D space with $2M$ lattice
points ($M$ points in each wire), corresponding either to real atoms
in a 1D chain or to a districtized continuous 1D space. Introducing
interactions, the Hamiltonian for this system is
$\hat{H} = \hat{H}_u + \hat{H}_l + \hat{H}_\mu + \HIWI + \HIWT$
with individual terms
\begin{multline*}
\hat{H}_{w} = - \sum_{x=1}^M t_w 
	\left( \hat{c}_{xw}^\dagger \hat{c}_{x+1w} + \mbox{h.c} \right) \\
	+ \sum_{x,y=1}^{M}U_w(|x-y|)\hat{n}_{xw}\hat{n}_{yw},
\end{multline*}
and
\begin{equation*}
\hat{H}_{\mu} = \sum_{x=1}^M 
	\frac{\mudiff}{2}( \hat{n}_{xu} - \hat{n}_{xl} ).
\end{equation*}
Here, $w\in \{u,l\}$ is the wire index, $\hat{c}_{xw}$ and
$\hat{c}_{xw}^\dagger$ are electron field annihilators and creators
at site $x$ in wire $w$,
$\hat{n}_{xw}=\hat{c}^\dagger_{xw}\hat{c}_{xw}$,
and $U_u=U_l$ is intrawire electron--electron interaction strength.
The opposite band curvatures imply $t_u=-t_l \equiv t>0$, and the
chemical potential difference $\mudiff$ is used to tune the filling fraction
of electrons inside each wire.
The IW terms are
\begin{gather*}
	\HIWI  = \sum_{x,y=1}^M U_{ul}(|x-y|)\hat{n}_{xu}\hat{n}_{yl}, \\
	\HIWT   = -t_\perp\sum_{x=1}^M 
		\left(\hat{c}_{xu}^\dagger \hat{c}_{xl} +\mbox{h.c}\right).
\end{gather*}
where $U_{ul}$ is the IW interaction potential. 
To simplify the analysis and keep the required computational effort
under control, we treat \textit{spinless} electrons, as
could be achieved, for example, by external magnetic
fields (see Supplementary Materials).

\begin{figure}[t]
\centering
\includegraphics[width=1\columnwidth,trim = 58mm 0mm 68mm 5mm, clip]
{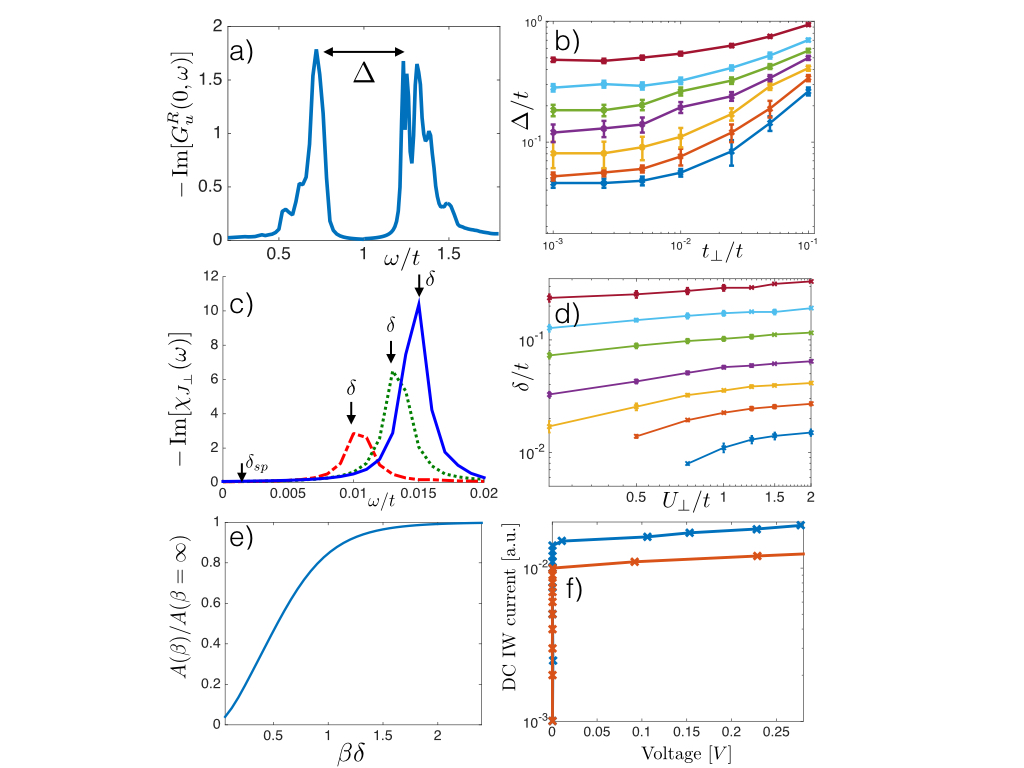}
\caption{
(a) Spectral function of $G_u^R(x,\omega)$  for the model system
with $t_\perp=0.001t$ and $U_\perp=2t$, exhibiting the large gap
$\Delta$.
(b) Scaling of $\Delta$ with $t_\perp$ for $U_\perp=0.25t$ (dark
blue), $U_\perp=0.5t$ (bright red), $U_\perp=0.75t$ (yellow), $U_\perp=t$
(violet) $U_\perp=1.25t$ (green), $U_\perp=1.5t$ (light blue),
$U_\perp=2t$ (dark red) for the model system.  
(c) Spectral function of  $\chi_{J_\perp}(\omega)$ for the model system
with $t_\perp=0.001t$ and $\eta=0.001t$, for 
$U_\perp=2t$ (blue line), 
$U_\perp=1.5t$ (green dotted), and
$U_\perp=t$ (red dash-dotted).  
Weight below $\delta$ is entirely due to finite $\eta$. 
(d) Scaling of $\delta$ with $U_\perp$, for $t_\perp=0.001t$ (dark
blue), $t_\perp=0.0025t$ (bright red), $t_\perp=0.005t$ (yellow),
$t_\perp=0.01t$ (violet) $t_\perp=0.025t$ (green), $t_\perp=0.05t$
(light blue), $t_\perp=0.1t$ (dark red) for the model system.  
(e) Order parameter $A$ as a fraction of its ground state value, against
inverse temperature $\beta$ for the model system with $U_\perp=2t$,
$t_\perp=0.01t$. Once $\beta>1/\delta$, the system approaches ground
state properties exponentially fast in $\beta$.
(f) DC I--V characteristic of the model system with $t=1\eV$,
$U_\perp=2t$, $t_\perp=0.01t$ (blue) and $t_\perp=0.001t$ (red), showing
both dissipationless and dissipative regimes. All results are for
$M=96$.}
\label{fig:Fig2}
\end{figure}

It is the IW tunneling that enables exciton condensation in 1D. 
A particle--hole transformation for the hole wire shows $\HIWT$ serving
as bias field for electron--hole pairs. 
Without interactions, the ODLRO due to $\HIWT$ is a trivial
single particle effect brought on by the opening of the single particle
gap $\deltasp$. 
In the following, we show that the IW repulsion between electrons
$\HIWI$ will feed strongly into this tiny nucleus of ODLRO and lead to a
truly many body EC. This gives a massive enhancement of EC properties
such as the temperature below which the system is close to the EC ground
state, and the response to applying IW current and voltage. These
properties are not affected by the explicit symmetry breaking nature of
IW tunneling, which will attempt to fix the global condensate phase to a
particular value.
This phase locking effect was studied for bilayer systems, and most
properties of the EC (including the technologically interesting ones)
only rely on having a large condensate amplitude~\cite{Su2008}.
Treating the interplay of IW tunneling and electron interactions in 1D
requires DMRG to fully capture the effects of nonperturbative $U_u$,
$U_l$, and $U_{ul}$. 

To illustrate the key features of the EC in 1D, we first study a model
system where electrons have no intrawire and purely local IW repulsion,
so that $U_{u}=U_{l}=0$ and $U_{ul}(|x-y|)=U_\perp\delta_{x,y}$. 
Once this is established, we show that with strong and
long range intrawire interactions, a nontrivial and measurable EC still
forms.

For the model system we calculate the ground states of $\hat{H}$ and
their exciton correlations $\Cex(x) = \langle
\hat{c}^\dagger_{0u} \hat{c}_{0l}\hat{c}^\dagger_{xl}\hat{c}_{xu}
\rangle$ for a grid of values of $t_\perp$ and $U_\perp$, fixing
the filling fraction in the electron wire at $0.1$. 
In Fig.~\ref{fig:Fig1}(c) we plot $\Cex(x)$ for $U_\perp=2t$.
The ODLRO is characterized by this exciton correlator approaching a
finite value at long distances. 
When $t_\perp=0$ this cannot happen and $\Cex(x)$ decays as 
$\propto x^{-K_a-1/K_s}$, as predicted by bosonization and MW (see the
Supplementary Material, and Ref.~\onlinecite{Werman2015}).
In contrast, when $t_\perp \neq 0$ the exciton correlator remains finite
at large $x$, indicating the presence of ODLRO and a stable EC.
Decreasing $t_\perp$ by an order of magnitude only halves the strength
of the ODLRO.
Because DMRG uses a lattice with open boundaries, we see end
effects where $\Cex(x)$ oscillates on a scale inversely
proportional to the small EC gap, $\delta$, described below. 
This is analogous to the penetration length of a superconductor. 
The noninteracting case, $U_\perp=0$, shown at $t_\perp\neq 0$ in
Fig.~\ref{fig:Fig1}(c) reveals the crucial importance of the IW
interactions for enhancing the magnitude of the ODLRO in the EC. 

To quantify directly how electron interactions dominate the EC
physics compared to the trivial gapped state of free electrons,
Fig.~\ref{fig:Fig1}(d) shows how the real space order parameter $A =
\langle\hat{c}^\dagger_{0u} \hat{c}_{0l}\rangle$ of
the EC is boosted over the corresponding value for $\HIWI=0$, which 
is set entirely by $t_\perp$.  
This order parameter also quantifies the ODLRO, since $\Cex(x)
\rightarrow A^2$ when $x\rightarrow\infty$.

Experimental observables capture how the IW interaction $U_\perp$
dominates the 1D EC physics.
Fundamentally, the 1D EC is not characterised by one gap, but by two,
which we label $\delta$ and $\Delta$. 
The large gap $\Delta$ could be measured using scanning tunneling
microscopy, which probes the retarded Green's function 
\begin{multline*} \label{stmresp}
	G_w^R(x,\omega) = 
		\langle \hat{c}_{xw} (\omega-\hat{H}+E_{GS}+i\eta)^{-1}
		\hat{c}^\dagger_{xw}\rangle \\
	+ \langle \hat{c}^\dagger_{xw}(\omega+\hat{H}-E_{GS}+i\eta)^{-1}
		\hat{c}_{xw}\rangle.
\end{multline*}
An example is shown in Fig.~\ref{fig:Fig2}(a).
Weak coupling perturbative renormalization group (pRG) predicts
$\Delta\propto U_\perp^{1/(2-2K_a)}$ (see the Supplementary
Material and Ref.~\onlinecite{Werman2015}), and is tied to the appearance of
Coulomb drag~\cite{Nazarov1998,Klesse2000,Chou2015}.
Numerically we find pRG to be of limited validity, with $\Delta(U_\perp)$ actually
interpolating between (at least) two power laws in $U_\perp$, where the 
position of the crossover region depends on $t_\perp$
(see the Supplementary Material).

Crucially, DMRG reveals the dependence of $\Delta$ on $t_\perp$ (which
pRG cannot), shown in Fig.~\ref{fig:Fig2}(b). Two regimes of the
1D EC can be identified. At very small $t_\perp/t$, $\Delta$ is almost
independent of $t_\perp$. Here the physics is almost completely
dominated by electron--hole interactions and this is the cleanest form
of a 1D many body EC. 
The other regime, when $t_\perp/t>0.005$, has a significant dependence
of $\Delta$ on $t_\perp$ and a noticeable decrease of the order parameter
ratio in Fig.~\ref{fig:Fig1}(d), although that ratio still remains large
if $U_\perp/t$ is large.

The large gap $\Delta$ is present even when $t_\perp=0$ and there is no EC. 
The small gap $\delta$ behaves differently. This gap can be
obtained from the first peak  in the imaginary part of the IW current
susceptibility
\begin{equation*}\label{jresp}
	\chi_{J_\perp}(\omega) =
	\langle \hat{J}_\perp (\omega-\hat{H}+E_{GS}+i\eta)^{-1}
	\hat{J}_\perp \rangle,
\end{equation*}
which is accessible via optical conductivity measurements and is shown
in Fig.~\ref{fig:Fig2}(c). Here, $\hat{J}_\perp = \frac{i}{M}
\sum_{x=1}^M \left( \hat{c}^\dagger_{xu}\hat{c}_{xl}-{\rm h.c.} \right)$
is the discretized operator for IW current.  
We can also find $\delta$ by computing the first excited state above the
ground state within the same quantum number
sector~\cite{Schollwock2011} and this gives matching values.
This gap only appears when $t_\perp\neq 0$ and is key for establishing
the EC. 
The pRG predicts $\delta\propto t_\perp^{2/(4-K_a+K_s^{-1})}$ at weak
$t_\perp$, and a locking of the phase of the symmetric mode (see the
Supplementary Material and Ref.~\onlinecite{Werman2015}). 
However, pRG cannot characterize the order when both $t_\perp$ and
$U_\perp$ flow to strong coupling, or when the system starts out at
strong coupling.
The limitations of pRG are illustrated again by our finding that
$\delta(t_\perp)$ is not a pure power law, but consists of two such laws
which cross over into each other (see the Supplementary Material).
Thus, we have used DMRG to establish that the ordered phase of this
system has excitonic ODLRO and the mutual enhancement of IW tunneling
and interactions 
which pRG cannot deliver.
The numerics further reveal that the EC order parameter $A\propto
|\operatorname{Im}[\chi_{J_\perp}(\delta)]|^\gamma$ once $U_\perp$
becomes the dominant energy scale, where $\gamma$ is independent of
$t_\perp$ (see the Supplementary Material). 

The gap $\delta$ sets the temperature below which the 1D excitons will
be very close to the EC ground state, as shown by the condensate 
order parameter $A$ in Fig.~\ref{fig:Fig2}(e). Computed quasiexactly
using DMRG from the full thermal state $e^{-\beta \hat{H}}$ via the
purification approach~\cite{Schollwock2011}, $A(\beta)$ includes both
the energy  and entropy contributions to the free energy. We see
that, even though it is very weak, the IW tunneling explicitly
circumvents the standard argument of Landau and Lifshitz regarding the
impossibility of an ordered EC phase at finite temperature, and $A$ in
Fig.~\ref{fig:Fig2}(e) exhibits crossover behaviour. This is
analogous to the exact solution for the magnetization of a 1D Ising
chain: at zero external magnetic field, no magnetized phase is possible at finite
temperature, but any finite external field will give rise to a crossover
behaviour of magnetization with temperature~\cite{BookReichl2016},
exactly analogous to Fig.~\ref{fig:Fig2}(e) for $A(\beta)$.
 
We calculate $\chi_{J_\perp}(\omega)$ in the real frequency domain
(using the GMRES approach within DMRG~\cite{Kuhner1999}) on the isolated
system. With no external bath to dissipate energy, this approach cannot
obtain DC IW current in response to applying $\hat{J}_\perp$.
Still, for an isolated system the existence of nondissipative DC
interlayer supercurrent (which is the hallmark property for using the EC
state as a transistor~\cite{Banerjee-EDL30}) can be shown, as can
the transition to a dissipative regime beyond some critical current. Both
regimes are visible in Fig.~\ref{fig:Fig2}(f), which shows 
$\chi_{J_\perp}(0)$ as a function of voltage $V=2\pi\Gamma/e$, where
$\Gamma$ is the rate of macroscopic tunneling from the original to the
new ground state as $I\hat{J}_\perp$ is added to $\hat{H}$. 
We obtain $\Gamma$ from the decay of occupation from the original ground
state through calculation of the imaginary time Green's function
$\langle GS|e^{-\tau(\hat{H}+I\hat{J})}|GS\rangle\propto
e^{-\tau\Gamma}$ using time dependent DMRG. 
The result agrees very well with the qualitative prediction of the
singular relationship $I\propto -(\log V)^{-1}$.

\begin{figure}[t]
\centering
\includegraphics[width=1\columnwidth,trim = 95mm 82mm 108mm 90mm, clip]
{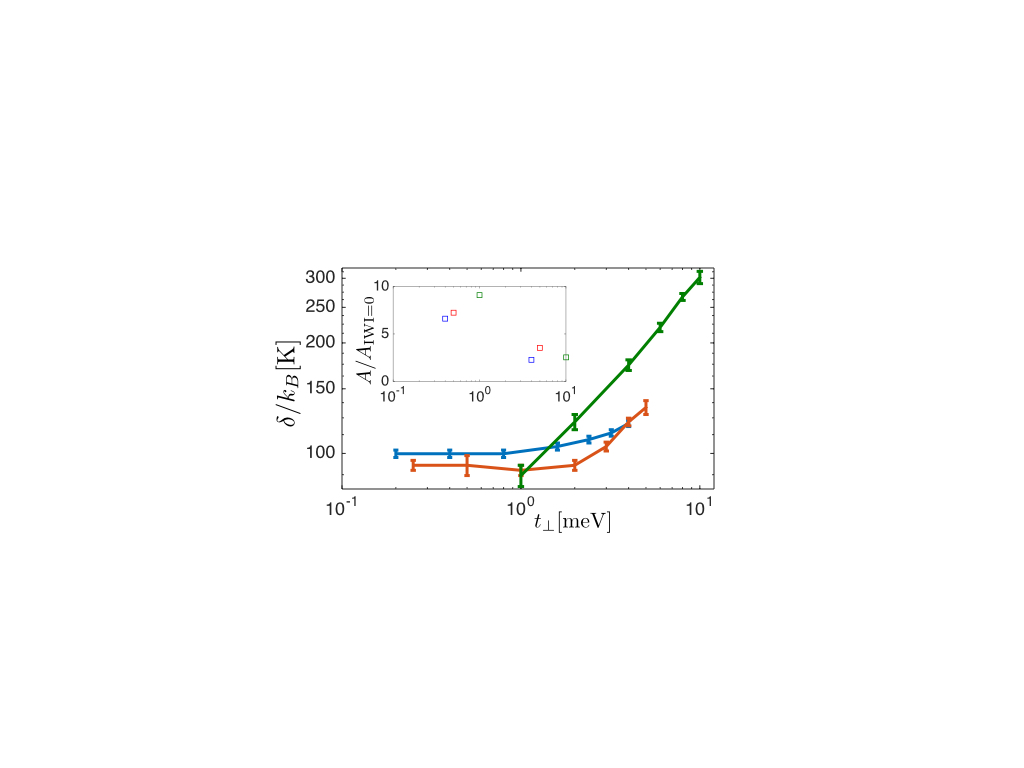}
\caption{Achievable gap $\delta/k_B$ in the strong
screening case as a function of $t_\perp$ for 
$t=1\eV$ (green line), 
$t=0.5\eV$ (red line), and
$t=0.4\eV$ (blue line).
Inset: range of achievable $A/A_{\rm{IWI}=0}$, indicated by showing high
and low values as a function of $t_\perp$, with colors matching the main
figure.}
\label{fig:Fig3}
\end{figure}

Realistic experimental systems have long range Coulomb interactions, so
now we demonstrate the robustness of the EC beyond the model system we
have considered up to this point.
We use a screened Coulomb potential (3D coordinates $\bf{x}$),
\begin{equation}\label{eq:uint}
	U(|\vx-\vx|)=
	\frac{e^{-|\vx-\vx'|/\Lambda}}
	{4\pi\varepsilon_{\mathrm{eff}}|\vx-\vx'|},
\end{equation}
where $\varepsilon_{\mathrm{eff}}$ denotes the effective dielectric
screening in between points $\vx$ and $\vx'$.
We note that for 1D electrons there is no \textit{intrinsic}
screening, a crucial advantage of implementing 1D excitons compared
to previous proposals in 2D bilayers~\cite{Sodemann-PRB85}. 
All screening in 1D derives from the
environment~\cite{Schulz1993a,Meyer2009} and can thus be tuned.
As shown in Fig.~\ref{fig:Fig1}(a), the dielectric constant of the
substrate $\epssub$ could be different from that of the spacer $\epssp$
if different materials are chosen.
For IW interactions $\varepsilon_{\mathrm{eff}}=\epssp$ and for
intrawire interactions
$\varepsilon_{\mathrm{eff}}=(\varepsilon_{\mathrm{sub}}+\epssp)/2$.
The aim is to depress intrawire repulsion as much as possible through
large $\varepsilon_{\mathrm{sub}}$, while retaining strong IW repulsion
through low $\epssp$. 
The particular form of the screening function in Eq.~\eqref{eq:uint} 
is secondary: what matters for is to choose a screening that (i) limits
the Coulomb interaction and (ii) reproduces the low energy properties of
a realistic wire.

With this in mind, we consider a lattice model of electrons on two
parallel chains, each with a lattice spacing $0.142\nm$ equal to the 
carbon--carbon bond of graphene and study two scenarios. 
(i) Moderate screening. 
Choosing $\epssub=16\varepsilon_0$ (where $\varepsilon_0$ is the vacuum
permittivity), $\Lambda=0.48\nm$, and $t=0.25\eV$, as explained in (Supplementary Material),
a single such wire
realizes system of strongly
correlated spinless electrons at a magnetic field of $0.06\T$. Its low energy properties are characterized by a
Tomonaga--Luttinger liquid
parameter~\cite{BookGiamarchi2003} $K=0.66$ (the model system had
$K=1$), which is comparable to some experimentally available nanowires.
We place two such wires $\dIW=1\nm$ apart with $\epssp=\varepsilon_0$. 
Taking $t_\perp = 0.25\meV$, we use DMRG to compute the ground state and
find that it exhibits ODLRO in $\mathcal{C}_\mathrm{ex}$, that
$\chi_{J_\perp}$ yields a substantial $\delta/k_B=88\K$,
and that $A/A_{{\rm IWI}=0}=5.4$. 
For temperatures below $\delta$, this system will be very close
to the EC ground state, having all the 1D EC properties established
earlier. 
(ii) Strong screening. 
Choosing $\epssub=200\varepsilon_0$, $\Lambda=0.31\nm$, a
wire is much closer to our initial model system than in the moderate
screening case. 
When $\dIW=0.75\nm$ and $\epssp=\varepsilon_0$, the IW interaction is
larger than intrawire one. We consider systems at $t=0.4,0.5,1.0\eV$, which can again be made spinless at perpendicular magnetic fields of $9.4\T$, $12\T$, and $27.9\T$ respectively (see Supplementary Materials).
The IW-tunneling $t_\perp$ is varied between $5\times10^{-4}t$
and $0.01t$. 
As shown in Fig.~\ref{fig:Fig3}, this results in $\delta/k_B$ of at least
$87\K$, reaching up to $301\K$. 
As for the model system earlier, we encounter both the regimes where
$\delta$ depends weakly on $t_\perp$, crossing over into one where
single particle physics becomes more noticeable. In the inset of
Fig.~\ref{fig:Fig3} we indicate the range of the corresponding
$A/A_{{\rm IWI}=0}$ by showing one high and one low value for each $t$.
For small values of $t_\perp$ we see the many body physics being 
clearly dominant. We find that $U_{u}$ and $U_l$ should not
be too different from $U_{ul}$ in overall magnitude. If intrawire
interactions are much stronger than IW ones, all EC
properties are depressed. In the opposite regime, where IW interactions
dominate, the electrons phase separate at large
$\mudiff$. 
As a result, we find different minimal filling fractions for the
electrons in the upper wire (low filling aids pairing) 
For (i) this is $0.135$, in (ii) it is $0.27$ at $t=0.4\eV$, $0.208$ at
$t=0.5\eV$, and $0.113$ at $t=1\eV$.

In conclusion, we have demonstrated that single particle tunneling
between spinless electron and hole wires removes the restrictions of MW
and allows for a true bilayer EC in one dimension characterized
by ODLRO and a global phase. 

Calculations were performed using the redeveloped DMRG module of ALPS
~\cite{Dolfi2014a} and the Matrix Product Toolkit~\cite{mptoolkit}. 
We thank Nordita for support.
A.K.~thanks Thierry Giamarchi for helpful discussions.
D.S.L.A.~thanks ERC project DM-321031 for financial support.

\bibliography{biblio}

\section{Supplementary Material}
\subsection{Perturbative RG of the bosonized model}
Though treating $\HIWI$ and $\HIWT$ perturbatively will not address
physically relevant systems, together with the associated bosonization
framework~\cite{BookGiamarchi2003} it does have some use for interpreting the
numerical results at strong coupling. 
After a particle--hole transformation $\hat{c}_{xl}\rightarrow
(-1)^x\hat{c}_{xl}^\dagger$ on $\hat{H}_l$, we bosonize
$\hat{H}_u+\hat{H}_l$, by retaining only the long wavelength excitations
around the Fermi points $\pm k_F$ in both wires, approximating
$\hat{c}_{xw}\propto U_Re^{i(k_Fx-\hat{\phi}_w(x)+\hat{\theta}_w(x))} +
U_Le^{-i(k_Fx-\hat{\phi}_w(x)-\hat{\theta}_w(x))}$, where
$\hat{\phi}_w(x)$, $\partial_x\hat{\theta}_{w}(x)$ are conjugate field
operators, $[\hat{\phi}_w(x),\partial_x\hat{\theta}_{w'}(x')] =
i\pi\delta_{ww'}\delta(x-x')$ associated with long wavelength density
and phase fluctuations in wire $w$ respectively, and $U_{R}$ and $U_{L}$
are the Klein factors that preserve anticommutation relations.
Thus, $\hat{H}_{w}$ becomes quadratic in
$\partial_x\hat{\phi}_{w}(x)$, $\partial_x\hat{\theta}_{w}(x)$ and
its long wavelength properties are parametrised by just two numbers,
the Tomonaga-Luttinger liquid (TLL) parameters $v_{w}$ and $K_{w}$.
The TLL parameters $K_{w}$ encode the strength and range of $U_{w}$
respectively and if $U_{w}=0$, then $K_{w}=1$.  The stronger and
more long ranged $U_{w}$ is, the further below $1$ the value of
$K_{w}$ will drop. 

Since we assume $|t_u|=|t_l|$, $U_u=U_l$, and $k_F$ being the same for
both wires, we have $K_u=K_l=K$ in the following.  Now adding
$\HIWI$ as perturbation to $\hat{H}_u+\hat{H}_l$, its bosonized
form in momentum space decomposes into a forward scattering part, with
terms proportional to $U^{F}_{ul}=U_{ul}(q=0)$ and a backscattering
contribution proportional to $U^{B}_{ul}=U_{ul}(q=2k_F)$.
The forward scattering term can be incorporated into
$\hat{H}_u+\hat{H}_l$ exactly, at the price of a canonical
transformation to the symmetric and antisymmetric modes of the two wires
$\hat{\phi}_{s,a} =(\hat{\phi}_u\pm\hat{\phi}_l)/\sqrt{2}$,
$\hat{\theta}_{s,a} =(\hat{\theta}_u\pm\hat{\theta}_l)/\sqrt{2}$.  This
results in $\hat{H}_u+\hat{H}_l\rightarrow\hat{H}_s+\hat{H}_a$, where
$\hat{H}_{s}$ and $\hat{H}_{a}$ are TLL Hamiltonians, and in perturbation theory
their TLL parameters are $K_{s,a}=((K)^{-2}\mp U^{F}_{ul}m_u/(2\hbar
k_FK))^{-1/2}$.  However, the backscattering part can at best be treated
using the perturbative renormalization group (pRG), and the same holds
for the bosonized version of $\HIWT$. On its own, the IW backscattering
(when relevant) is the source of Coulomb drag
\cite{Nazarov1998,Klesse2000,Chou2015}. 

As in Ref.~\cite{Werman2015}, treating both perturbations jointly in
second-order momentum space pRG gives
\begin{equation}\label{rgeq}
	\frac{dU^{B}_{ul}}{ds}=2(1-K_a)U^{B}_{ul}, \quad
	\frac{dt_\perp}{ds}=\frac{\left(4-K_a-K_s^{-1}\right)t_\perp}{2}.
\end{equation} 
Thus, both $\HIWI$ and $\HIWT$ are relevant perturbations for a very
wide range of parameters (for example, $\HIWI$ is so for any repulsive
$U_{ul}$), and their associated couplings both flow to nonperturbative
values, outside the range of any pRG. The validity of Eq.~(\ref{rgeq})
is constrained further because the $U^{B}_{ul}$ may flow to its
fixed point before $t_\perp$, locking
$\hat{\phi}_a$ to a fixed value and making the pRG equation for
$t_\perp$ obsolete. If, against these objections, a straight
extrapolation of Eq.~(\ref{rgeq}) is performed, it would predict one
gap $\Delta$ for fluctuations of $\hat{\phi}_a$, with scaling
$\Delta\sim (U^{B}_{ul})^{1/2(1-K_a)}$, and another, $\delta$,
for fluctuations of $\hat{\theta}_s$, with scaling $\delta\sim
t_\perp^{2/(4-K_a+K_s^{-1})}$, the most relevant bosonized operators inside
$\HIWI$ and $\HIWT$ being compatible.
\subsection{Comparison to DMRG -- beyond perturbative RG}
\begin{figure}[t]
\centering
\includegraphics[width=1\columnwidth,trim = 85mm 0mm 85mm 0mm, clip]%
{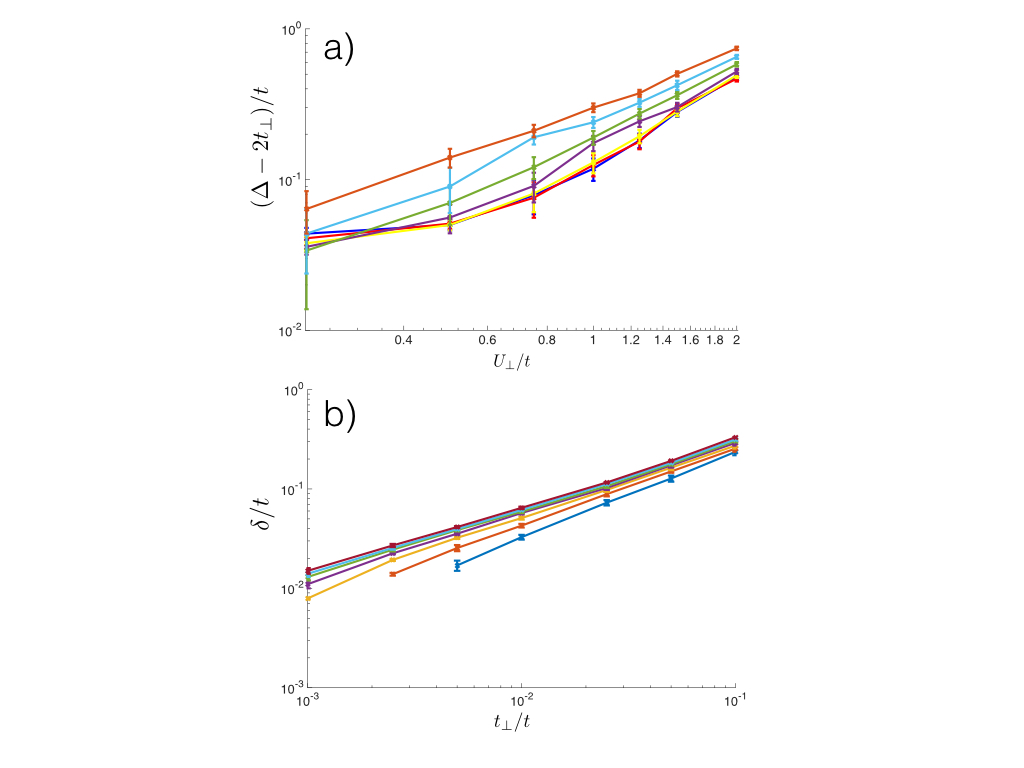}
\caption{
{\bf(a)} Interaction-dependent part of the gap $\Delta$, 
$\Delta-2t_\perp$,
plotted against $U_\perp$ for the ideal model system,
at $t_\perp=0.001t$ (dark blue), $t_\perp=0.0025t$ (light red),
$t_\perp=0.005t$ (yellow), $t_\perp=0.01t$ (violet), 
$t_\perp=0.025t$ (green), $t_\perp=0.05t$ (light blue),
$t_\perp=0.1t$ (dark red).
{\bf(b)} Gap $\delta$ against $t_\perp$ for the ideal model system,
at $U_\perp=0.25t$ (dark blue), $U_\perp=0.5t$ (light red),
$U_\perp=0.75t$ (yellow), $U_\perp=t$ (violet), 
$U_\perp=1.25t$ (green), $U_\perp=1.5t$ (light blue),
$U_\perp=2t$ (dark red).
}
\label{fig:scaling}
\end{figure}

With DMRG numerics we can address the regimes in which pRG is not valid.
The main text focuses on the most relevant effects beyond pRG, such as
the establishment of excitonic ODLRO, the mutual reinforcement of IW
interaction and IW tunneling leading
to enhancements of both $\delta$ and $A$, as well as computing $\delta$ quantitatively
for quasirealistic systems, 

We compare DMRG to the predictions that pRG can be used for, the scaling
of $\Delta$ with $U^{B}_{ul}$, and of $\delta$ with $t_\perp$ for our
ideal model system with $U_u=U_l=0$, $U_{ul}^B=U_\perp$. 
As can be seen in Fig.~\ref{fig:scaling}(a), $\Delta$ is governed by at
least two different power laws (as opposed to the single power law
predicted by pRG) and the position of the crossover between the two
power laws strongly depends on $t_\perp$. Note that we have subtracted
the trivial contribution to the gap $\Delta$ which comes from the band
hybridization and is equal to $2t_\perp$.

For the gap $\delta$ we find qualitatively similar results, as shown
in Fig.~\ref{fig:scaling}(b), but we only enter the crossover zone for
the lowest values of $U_\perp$ and $t_\perp$ within our parameter grid.

\subsection{Linking $A$ to observables}
\begin{figure}[t]
\centering
\includegraphics[width=1\columnwidth,trim = 45mm 35mm 45mm 35mm, clip]%
{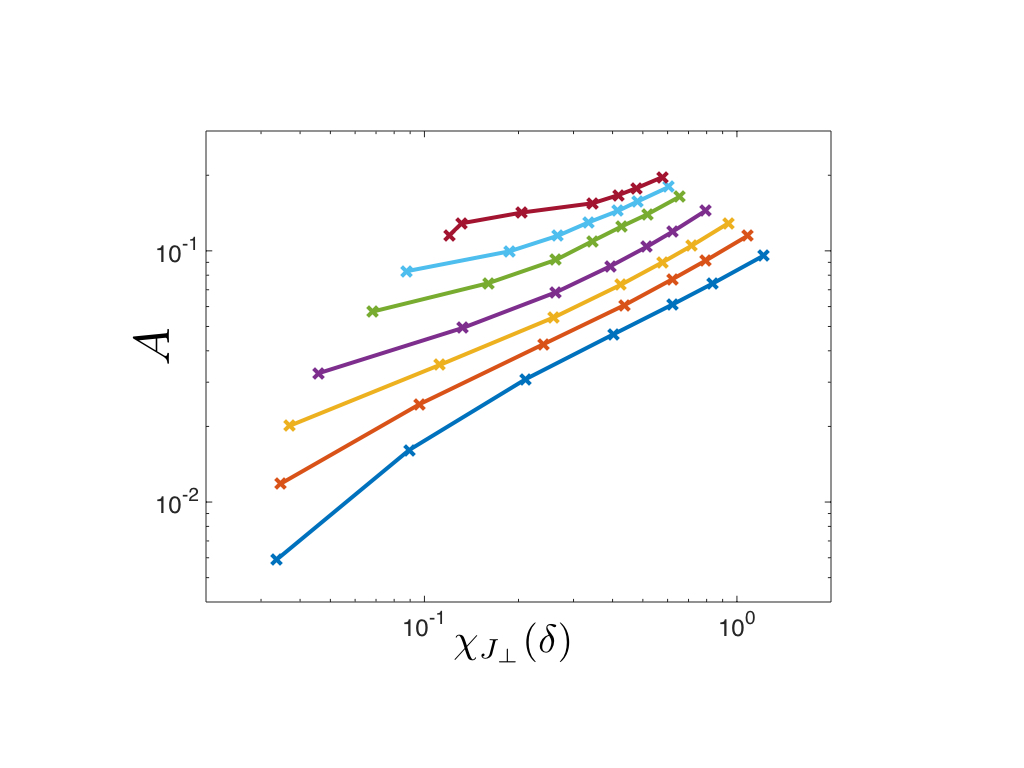}
\caption{
Order parameter $A$ as a function of the peak of the IW conductivity
$\chi_{J_\perp}$ at $\omega=\delta$, for $t_\perp=0.001t$ (dark blue),
$t_\perp=0.0025t$ (light red), $t_\perp=0.005t$ (yellow),
$t_\perp=0.01t$ (violet), $t_\perp=0.025t$ (green), $t_\perp=0.05t$
(light blue), $t_\perp=0.1t$ (dark red). All data taken at $\eta=0.01t$.
}
\label{fig:Aoverp}
\end{figure}
We show in the main text that the peak of the IW conductivity
$\chi_{J_\perp}(\omega)$ at $\omega=\delta$ is a proxy for the order
parameter $A$ of the 1D EC, and is the figure of merit for proposed
applications of bilayer ECs as low power electrical switches.
We show in Fig.~\ref{fig:Aoverp}, that as $\chi_{J_\perp}(\omega)$
increases with $U_\perp$ for
any given $t_\perp$ (the lines in Fig.~\ref{fig:Aoverp} correspond to
different values of $t_\perp$), the dependency of $A$ on
$\chi_{J_\perp}(\delta)$ seems to enter a power law scaling with an
exponent $\gamma$. Within the resolution of our grid of $\omega$ values,
the value of $\gamma$ for each $t_\perp$ appears to be either the same,
or so close to each other that we cannot resolve the difference.

\subsection{Obtaining a spin-polarized system}
For the quasi-realistic systems with long range interactions (Eq.~(1)
of the main text), the magnetic field needed to prepare the system
in the fully spin-polarized state that we study can be
straightforwardly computed using DMRG. We calculate the energy
of the ground state $E_{\rm GS}(N_\uparrow=N_\downarrow)$ in the
spin-balanced system, as well as for the spin-polarized system,
$E_{\rm GS}(2N_\downarrow)$. The long range
interactions are exactly the same in each case, but in our discretized
spinful wire we assume an onsite repulsion $U/t=10$. The magnetic Zeeman
energy of an electron in the spinful system with an external magnertic
field $B$ perpendicular to the wires is given by 
$E_{\mathrm{mag}}(\sigma) = g \mu_B B \sigma$, with $\sigma=\pm1/2$, 
the Lande $g$-factor $g=2$ and the Bohr magneton $\mu_B = 5.788\times
10^{-5}$ $\eV/\T$.  Thus, at
\begin{equation}
B_{\rm pol} =\frac{E_{\rm GS}(2N_\downarrow)-E_{\rm GS}(N_\uparrow,N_\downarrow)}{g\mu_BN_\downarrow}
\end{equation}
the magnetic field is strong enough to polarize both wires. 

Thus, in the non-polarized case at $B=0$, one would start with the
spin-degenerate electron-like bands of the upper wire overlapping with
the hole-like bands
of the lower wire, but at the same time the Fermi level would be
below the band minimum of the upper wire. Generically, it is possible to
achieve this by employing back gates.
Then, as $B$ becomes non-zero, the spin-down bands of
both wires will be shifted down in energy 
until the Fermi level lies in the hybridization gap of the spin-down
bands. This is the situation depicted in Fig. 1b of the main text, and
as long as $B>B_{\rm pol}$ holds, this system is fully spin-polarized.

%\bibliography{biblio}

\end{document}